# HORIZONTAL-COMPONENT PRIOR-BASED FRAMEWORK FOR ADAPTIVE SHEAR-WAVE LEAKAGE SUPPRESSION IN OBC DATA

Authors — Zheng Cong, Shiqi Dong, Xintong Dong and Xunqian Tong.

Zheng Cong, State Key Laboratory of Deep Earth Exploration and Imaging, College of Instrumentation and Electrical Engineering, Jilin University, Changchun City, Jilin. E-mail: cz19990714@126.com.

Shiqi Dong, Key Laboratory of Modern Power System Simulation and the Control and Renewable Energy Technology, Ministry of Education, Jilin City, Jilin, China, and Department of Communication Engineering, College of Electric Engineering, Northeast Electric Power University, Jilin City, Jilin. E-mail: dsq1994@126.com.

Xintong Dong, State Key Laboratory of Deep Earth Exploration and Imaging, College of Instrumentation and Electrical Engineering, Jilin University, Changchun City, Jilin. E-mail: 18186829038@163.com; ORCID: 0000-0002-6887-8712.

Xunqian Tong, Key Laboratory of Deep Earth Exploration and Imaging, College of Instrumentation and Electrical Engineering, Jilin University, Changchun City, Jilin. E-mail: txq@jlu.edu.cn.

Corresponding author: Xintong Dong, 18186829038@163.com

Three sentence novelty statement: In this work, we propose a horizontal-component prior-based framework (HPAS) that enables adaptive shear-wave leakage suppression in OBC data without requiring clean labels, addressing a key limitation of conventional supervised deep-learning approaches. By introducing an additive–subtractive noise construction strategy with moment matching, the method generates statistically consistent training pairs directly from raw field data and establishes a theoretical link to supervised learning through loss-function analysis. The proposed framework improves both noise attenuation and signal preservation in complex cases, offering a new viable approach for shear-wave suppression.

Running head: Prior guided S-wave leakage suppression

# ABSTRACT

Shear-wave leakage in the vertical (Z) component of ocean-bottom cable (OBC) seismic data commonly results from the receiver tilt and poor seafloor coupling, introducing unwanted coherent noise that impacts the subsequent data processing and imaging. Traditional denoising methods are limited by manual parameter tuning and idealized model assumptions, while deep-learning (DL) approaches have shown significant potential in suppressing shear-wave leakage. However, supervised learning requires clean primary waves (P waves) as the label, which is generally impractical to obtain for field data. To address these challenges, we propose a framework based on horizontal-component priors for adaptive shear-wave leakage suppression (HPAS). Instead of relying on clean primary-wave (P-wave) data, HPAS generates input-label pairs directly from raw multi-component field data using an additive-subtractive noise strategy. Specifically, we extract shear-wave (S-wave) noise from the horizontal components and apply a linear transformation to match its first and second order moments with the S-wave leakage in the Z-component, and the statistically matched noise is then added to and subtracted from the original Z-component to create the input and label pairs. By allowing the denoising model to learn the S-wave features present in the differences between the input and the label, the adaptive denoising process approximates supervised learning. Evaluations on both synthetic and field data demonstrate that the proposed HPAS framework effectively and adaptively suppresses S-wave leakage while preserving the amplitude of the P-wave signals in the Z-component, offering a robust solution with strong generalization capabilities.

## INTRODUCTION

Ocean Bottom Cable (OBC) acquisition is often used to obtain high-quality multi-component seismic data (Caldwell, 1999; Stewart et al., 2003). Ideally, the vertical (Z) component mainly records primary waves (P waves). However, due to the receiver tilt and poor seafloor coupling, a portion of the shear waves (S waves) leaks into the Z-component of receivers (Jolly, 1956; Wang and Wang, 2017), which can be regarded as a type of coherent noise. Therefore, methods aimed at mitigating S-wave leakage while preserving the P waves are essential. The conventional methods for mitigating S-wave leakage can be divided into three categories: τ–p transform–based methods, wavefield rotation–based methods, and transform-domain adaptive filtering methods.

In early studies, researchers proposed a series of methods based on the time-intercept-slowness (τ-p) transformation (Durrani and Bisset, 1984; Kappus et al., 1990) to separate S waves from Z-components. Tatham and Goolsbee (1984) transformed seismic data from the time-space (t-x) domain to the τ-p domain, mitigating the S-wave leakage based on the distinct moveout characteristics of P and S waves. Building upon this approach, Wang et al. (2002)

developed a wavefield-rotation method that rotates the horizontal and vertical components separately to align the P- and S-wave fields with their observed polarization directions, thereby achieving P- and S-wave separation. In addition to τ-p–based approaches, several alternative strategies have been proposed to address the S-wave leakage problem. Yu et al. (2011) exploited the differences in physical responses between hydrophones and geophones to develop a local attribute matching (LAM) filter based on the two-dimensional complex wavelet transform (CWT). While this method achieves adaptive denoising, its effectiveness relies on the assumption that the P-component is relatively clean. Thus, the performance of the LAM filter typically deteriorates in the presence of strong background noise. Yang et al. (2020) exploited the directional sparsity and edge-preserving properties of the curvelet transform to decompose seismic data into multiscale and multidirectional components, achieving simultaneous P-Z matching and noise attenuation. Although these conventional methods have a certain effect in suppressing the S-wave leakage, most of them rely on idealized prior assumptions, such as linearly structured events, travel-time contrasts, and noise-free conditions. Because these idealized conditions are difficult to be satisfied during field data acquisition, the performance of these methods often degrades when applied to complex data.

In recent years, deep learning (DL) has emerged as an effective tool for seismic data processing, owing to its capability to construct complex nonlinear relationships directly from data (Hinton et al., 2006; LeCun et al., 2015; Qiu et al., 2016; Wang et al., 2020). DL has been increasingly applied to various tasks, including seismic denoising (Dong et al., 2019; Iqbal, 2022), data interpolation (Wang et al., 2019; Dong et al., 2025), resolution enhancement (Zhong et al., 2024; Yu et al., 2025), first-break picking (Yuan et al., 2018; Yin et al., 2023), frequency extrapolation (Sun and Demanet, 2021; Dong et al., 2024), and seismic inversion (Chen and Saygin, 2021; Wang et al., 2023). DL approaches have also shown significant potential in S-wave leakage suppression. Sun et al. observed that the shear-wave leakage in common-receiver gathers (CRGs) manifests as coherent noise, which is suitable to remove using DL methods. They used the horizontal- and P-component CRG data as training noise and label for the training of U-Net, confirming that DL methods are capable of capturing the feature of shear-wave leakage. However, using field data directly as labels may be insufficiently accurate, potentially affecting the performance of the model. To address this issue, Wang et al. (2023) proposed a S-wave predicting model using clean synthetic P waves as labels and their combinations with field S waves as noisy Z-components, which dynamically separates S waves from Z-components during training to improve denoising performance. Chen et al. (2024) employed a DL model to map S-wave features from horizontal components to the S-wave leakage in the Z-component, effectively suppressing the S-wave leakage via self-supervised noise subtraction. However, DL models trained on synthetic data typically exhibit limited generalization to field data. Moreover, the prediction of S-wave leakage solely from raw horizontal components may fail to handle complex data, due to the feature differences between the horizontal components and S-wave leakage.

To address these challenges, we propose a framework based on horizontal-component priors for adaptive S-wave leakage suppression (HPAS) in OBC data. The proposed framework constructs training pairs from field OBC horizontal and vertical component data using an additive-subtractive noise strategy (Monroy et al., 2025). Specifically, S-wave features are first extracted from the horizontal components via the Radon transform. The S-wave noise used for training is then generated by matching the first- and second-order statistical moments of the estimated S-wave energy in the Z-component. This statistically consistent S-wave noise is added to and subtracted from the original Z-component to generate the input and label pairs, and then a DnCNN (Zhang et al., 2017) is applied as the noise prediction model. This strategy is theoretically supported by a Taylor expansion analysis of the loss function, where the optimization of the network parameters is statistically equivalent to a supervised learning process. Therefore, the network can predict S-wave leakage from noisy inputs without requiring P waves. Evaluations on both synthetic and field data demonstrate that the proposed HPAS framework substantially reduces S-wave leakage while preserving P-wave signals, outperforming conventional approaches and showing strong generalization capabilities.

# METHODS

This section presents the methodology of the proposed HPAS framework. We describe the generation strategy for the training dataset and the denoising principle of the HPAS framework, followed by the implementation details of the DnCNN training and testing.

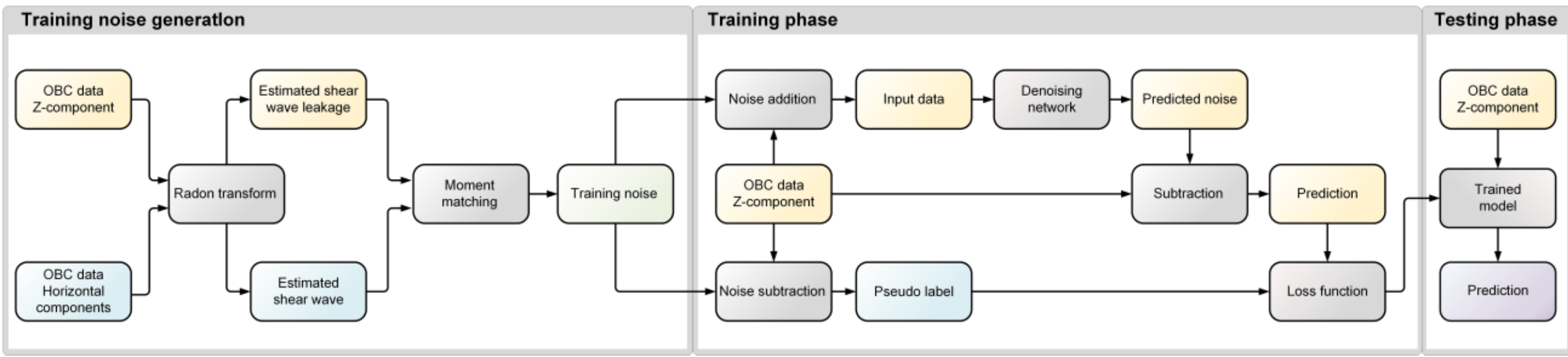

Figure 1. Horizontal-component priors-based framework for adaptive shear-wave leakage suppression.

## Denoising theory

The original Z-component data can be decomposed into P waves and S-wave leakage:

$$\mathbf{z} = \mathbf{x} + \mathbf{n}, \tag{1}$$

where $\mathbf{z}$, $\mathbf{x}$, and $\mathbf{n}$ represent noisy Z-component data, clean P-wave signal, and S-wave leakage noise, respectively. Monroy et al. (2025) proposed a noise prediction model that trains a neural network based on the constructed input and label pairs. The training data pairs are constructed as follows:

$$\mathbf{y}_1 = \mathbf{x} + \mathbf{n} + \mathbf{s},$$

$$\mathbf{y}_2 = \mathbf{x} + \mathbf{n} - \mathbf{s}, \tag{2}$$

where $\mathbf{y}_1$ and $\mathbf{y}_2$ represent the results of adding and subtracting noise to the noisy data, serve as the network input and pseudo-label, respectively. $\mathbf{s}$ represents the artificially added noise that is designed to be statistically similar to $\mathbf{n}$, and its generation will be described in detail later. The loss function $L$ used for training can be expressed as:

$$\begin{aligned} L &= E\left\|\mathrm{f}(\mathbf{y}_1) - \mathbf{y}_2\right\|_2^2 \\ &= E\left\|(\mathrm{f}(\mathbf{y}_1) - \mathbf{x}) + (-\mathbf{n} + \mathbf{s})\right\|_2^2 \\ &= E\left\|(\mathrm{f}(\mathbf{y}_1) - \mathbf{x})\right\|_2^2 + 2E\left\langle \mathrm{f}(\mathbf{y}_1) - \mathbf{x}, -\mathbf{n} + \mathbf{s}\right\rangle + E\left\|(-\mathbf{n} + \mathbf{s})\right\|_2^2, \end{aligned} \tag{3}$$

where $\mathrm{f}$ denotes the mapping relationship fitted by the model, and $E$ represents the expectation. $E\langle \bullet, \bullet \rangle$ denotes inner product over all samples. $E\left\|(-\mathbf{n} + \mathbf{s})\right\|_2^2$ is a term independent of the network parameters and therefore makes no contribution to the gradient during the training. We perform a Taylor expansion at $\mathbf{x}^i$ for each sample $\mathrm{f}^i(\mathbf{y}_1^i)$ of the network output $\mathrm{f}(\mathbf{y}_1)$, yielding:

$$\begin{aligned} \mathrm{f}^i(\mathbf{y}_1^i) &= \mathrm{f}^i(\mathbf{x}^i + \mathbf{n}^i + \mathbf{s}^i) \\ &= \sum_{k=0}^{\infty} \frac{1}{k!} \frac{\partial^k \mathrm{f}^i}{\partial (\mathbf{x}^i)^k} (\mathbf{x}^i)(\mathbf{n}^i + \mathbf{s}^i)^k, \end{aligned} \tag{4}$$

where $k$ represents the order of the Taylor expansion, $i$ represents the index of each sample, which ranges from 1 to $q$, where $q$ is the total number of samples. By inserting the Taylor expansion at $\mathbf{x}^i$ of the network output $\mathrm{f}^i(\mathbf{x}^i + \mathbf{n}^i + \mathbf{s}^i)$ from equation 4 into the loss function shown in equation 3, we obtain:

$$L \propto E_{\mathbf{n},\mathbf{s}} \left\| (\mathrm{f}(\mathbf{y}_1) - \mathbf{x}) \right\|_2^2 + 2E \left\langle \sum_{i=1}^{n} \sum_{k=0}^{\infty} \frac{1}{k!} \frac{\partial^k \mathrm{f}^i}{\partial (\mathbf{x}^i)^k} (\mathbf{x}^i)(\mathbf{n}^i + \mathbf{s}^i)^k - \mathbf{x}^i, -\mathbf{n}^i + \mathbf{s}^i \right\rangle. \tag{5}$$

In this study, $\mathbf{n}$ and $\mathbf{s}$ represent the S-wave leakage in the Z-component and S waves extracted from the original horizontal components, respectively. Since $\mathbf{x}$ can be considered approximately independent of $\mathbf{n}$ and $\mathbf{s}$, the loss function can be approximated as:

$$L \propto E_{\mathbf{n},\mathbf{s}} \left\| (\mathrm{f}(\mathbf{y}_1) - \mathbf{x}) \right\|_2^2 - 2 \sum_{i=1}^{n} \sum_{k=0}^{\infty} \frac{1}{k!} \frac{\partial^k \mathrm{f}^i}{\partial (\mathbf{x}^i)^k} (\mathbf{x}^i) E (\mathbf{n}^i + \mathbf{s}^i)^k (\mathbf{n}^i - \mathbf{s}^i). \tag{6}$$

We consider the loss function in Equation 6 for the cases of $k$ = 0, 1, 2, separately:

$$L \propto E_{\mathbf{n},\mathbf{s}} \left\| (\mathrm{f}(\mathbf{y}_1) - \mathbf{x}) \right\|_2^2 - 2 \sum_{i=1}^{n} \mathrm{f}^i (\mathbf{x}^i) E (\mathbf{n}^i - \mathbf{s}^i), \ \textbf{when } k = 0,$$

$$L \propto E_{\mathbf{n},\mathbf{s}} \left\| (\mathrm{f}(\mathbf{y}_1) - \mathbf{x}) \right\|_2^2 - 2 \sum_{i=1}^{n} [\mathrm{f}^i (\mathbf{x}^i) E (\mathbf{n}^i - \mathbf{s}^i) + \frac{\partial \mathrm{f}_i}{\partial \mathbf{x}_i} (\mathbf{x}^i) E ((\mathbf{n}^i)^2 - (\mathbf{s}^i)^2)], \ \textbf{when } k = 1,$$

$$\begin{aligned} L \propto E_{\mathbf{n},\mathbf{s}} \left\| (\mathrm{f}(\mathbf{y}_1) - \mathbf{x}) \right\|_2^2 - 2 \sum_{i=1}^{n} [\mathrm{f}^i (\mathbf{x}^i) E (\mathbf{n}^i - \mathbf{s}^i) + \frac{\partial \mathrm{f}_i}{\partial \mathbf{x}_i} (\mathbf{x}^i) E ((\mathbf{n}^i)^2 - (\mathbf{s}^i)^2) \\ + \frac{1}{2} \frac{\partial^2 \mathrm{f}_i}{\partial \mathbf{x}_i^2} (\mathbf{x}^i) E ((\mathbf{n}^i)^3 + (\mathbf{n}^i)^2 \mathbf{s}^i - \mathbf{n}^i (\mathbf{s}^i)^2 - (\mathbf{s}^i)^3)], \ \textbf{when } k = 2. \end{aligned} \tag{7}$$

Ideally, when $\mathbf{n}$ and $\mathbf{s}$ are independent and have zero mean, the cross terms in the Taylor expansion of the loss function in Equation 7 can be regarded as zero. For the case of $k$ = 0, 1, 2, $\mathbf{n}$ and $\mathbf{s}$ satisfy the following conditions:

$$E\left[ (\mathbf{n}^i)^k \right] = E\left[ (\mathbf{s}^i)^k \right], \quad k = 0,1,2, \tag{8}$$

In this case, the loss function shown in Equation 7 can be approximated by the supervised loss function:

$$L \approx E\left\| (\mathrm{f}(\mathbf{y}_1) - \mathbf{x}) \right\|_2^2 . \tag{9}$$

In the S-wave suppression task, the S-wave leakage in the Z-component exhibits similar features to the S-wave in the horizontal components, corresponding to $\mathbf{n}$ and $\mathbf{s}$ within $\mathbf{y_1}$. The network learns the features of the S waves by minimizing the difference between $\mathbf{y_1}$ and $\mathbf{x}$, thereby obtaining the clean data by subtracting the predicted S waves from the noisy data $\mathbf{z}$. The workflow of the HPAS framework is illustrated in Figure 1. To ensure the validity of Equation 9, the high-order terms in the Taylor expansion given by Equation 7 must vanish, which necessitates that $\mathbf{n}$ and $\mathbf{s}$ satisfy the high-order moment matching condition. However, the original S waves $\mathbf{h}$ in the horizontal components do not inherently share the same moments as the S-wave leakage in the Z-component. Therefore, we apply a linear transformation to $\mathbf{h}$ to obtain the training noise $\mathbf{s}$ that matches the moment of the S-wave leakage. We use the Radon transform to estimate the S waves $\mathbf{n}$ in the Z-component and its moments of different orders, and subsequently apply a linear transformation to the S waves $\mathbf{h}$ in the horizontal components to generate the training noise $\mathbf{s}$. However, if a linear transformation is applied to $\mathbf{h}$ to match its high-order moments with those of $\mathbf{n}$, the transformed $\mathbf{h}$ (i.e. $\mathbf{s}$) will lose the S-wave features, which will affect the training results. Our goal is to preserve as many of the S-wave features in the training noise as possible for efficient network learning. In addition, ensuring the statistical independence between the S waves $\mathbf{n}$ in the Z-component and the training noise $\mathbf{s}$—a key assumption for higher-order terms—is difficult to achieve when using complex field data. Consequently, the higher-order terms in the Taylor expansion cannot be approximated as zero, which introduces additional bias into the loss function and leads to training errors. Therefore, we restrict our analysis to the case of $k = 1$. The choice of the approximation order directly determines the statistical conditions that must be satisfied. When $k = 1$, the bias terms in the loss function depend primarily on the expected values of the first- and second-order moments of the noise distribution. To enable the loss to effectively approximate the supervised loss, these moment-induced biases must be removed. Therefore, it is necessary to match only the first- and second-order moments between $\mathbf{s}$ and $\mathbf{n}$. The linear transformation corresponding to the matching of $\mathbf{s}$ and $\mathbf{n}$ can be expressed as:

$$\mathbf{s} = \sqrt{\frac{\sigma_{\mathbf{n}}^2}{\sigma_{\mathbf{h}}^2}}\left(\mathbf{h} - \mu(\mathbf{h})\right) + \mu(\mathbf{n}), \tag{10}$$

where $\sigma_{\mathbf{n}}$ and $\sigma_{\mathbf{h}}$ represent the variance of $\mathbf{n}$ and $\mathbf{h}$, respectively. $\mu_{\mathbf{n}}$ and $\mu_{\mathbf{h}}$ represent the mean of $\mathbf{n}$ and $\mathbf{h}$, respectively. In this case, both $\mu_{\mathbf{s}} = \mu_{\mathbf{n}}$ and $\sigma_{\mathbf{s}}^2 = \sigma_{\mathbf{n}}^2$ are satisfied, and the second term of the loss function in Equation 6 can be approximated as zero. After $\mathbf{s}$ is added to and subtracted from the Z-component, the paired input and pseudo-label are constructed for training. According to Equation 9, the network effectively learns the combined features of $\mathbf{s}$ and $\mathbf{n}$. After moment matching, $\mathbf{s}$ and $\mathbf{n}$ have similar features, which enables the network to adaptively extract the shear-wave leakage from the Z-component. Notably, the $\mathbf{s}$ derived from the first- and second-order moment matching linear transformation of the original $\mathbf{s}$ preserves most S-wave features, thereby enabling the network to learn the features of S-wave leakage in Z-components. Regarding the selection of the network, we employ DnCNN (Zhang et al., 2017) as the noise prediction model due to its simple architecture and efficient residual learning strategy. This network has been widely applied in various seismic denoising tasks, such as random noise suppression, surface-wave suppression, and background-noise suppression.

**Evaluation Metrics**

We use the signal-to-noise ratio (SNR) and the structure similarity index measure (SSIM) to evaluate the accuracy of the model. The SNR and SSIM can be calculated as follows:

$$\mathrm{SNR}(\mathbf{p},\hat{\mathbf{p}}) = 10\log_{10}\frac{\sum_{i=1}^{M}\sum_{j=1}^{R}\hat{\mathbf{p}}(i,j)^2}{\sum_{i=1}^{M}\sum_{j=1}^{R}(\mathbf{p}(i,j)-\hat{\mathbf{p}}(i,j))^2}, \tag{11}$$

$$\mathrm{SSIM}(\mathbf{p},\hat{\mathbf{p}}) = \frac{(2\mu_{\mathbf{p}}\mu_{\hat{\mathbf{p}}}+c_1)(2\mathrm{Cov}_{\mathbf{p},\hat{\mathbf{p}}}+c_2)}{(\mu_{\mathbf{p}}^2+\mu_{\hat{\mathbf{p}}}^2+c_1)(\sigma_{\mathbf{p}}^2+\sigma_{\hat{\mathbf{p}}}^2+c_2)}, \tag{12}$$

where $\hat{\mathbf{p}}$ and $\mathbf{p}$ represent the ground truth and prediction, respectively. $M$ and $R$ represent the number of shots and time samples, respectively. $i$ and $j$ represent the shot index and time-sample index, respectively. $\mu_{\bullet}$ and $\sigma_{\bullet}$ represent the mean and standard deviation operator, respectively. $\mathrm{Cov}_{\mathbf{p},\hat{\mathbf{p}}}$ represents the covariance between $\hat{\mathbf{p}}$ and $\mathbf{p}$. $c_1$ and $c_2$ are constants introduced to ensure that the denominator is nonzero. SNR reflects the strength of the useful signal relative to the noise. A higher SNR in the prediction indicates better signal recovery quality and lower noise energy. SSIM reflects the structural similarity between the predicted results and the reference data. An SSIM value closer to 1 indicates that the feature distribution in the prediction is more consistent with that of the reference data. Notably, since receiver-related features remain fixed within a common-receiver gather while source locations vary, S-wave events in common-receiver gathers (CRGs) exhibit better continuity of events than those in common-shot gathers (CSGs). This enhanced continuity makes it easier for DL methods to capture the features of the S waves from the CRGs. Therefore, both the training and testing of the network are based on CRGs in our study.

### Experimental environment and training parameters

Hardware equipment significantly affects the efficiency of DL methods. In this study, the experimental environment consists of a CPU (Intel® Core™ i5-14600KF processor) and an NVIDIA GeForce RTX 4060 (8 GB VRAM). The network implementation is built on PyTorch (version 2.0.0). As shown in Table 1, the batch size is set to 1, the initial learning rate is 0.001, and the learning rate decreases by a factor of 0.3 every 50 epochs. The total number of training epochs is 200. To improve the stability of network training, the Kaiming initialization method is adopted (He et al., 2015).

TABLE 1

| Parameters | HPAS |
|---|---|
| Optimizer | Adam |
| Batch size | 1 |
| Epoch number | 200 |
| Learning rate range | $[10^{-3},10^{-5}]$ |
| Learning rate change interval | 40 |

## NUMERICAL EXAMPLES

In this section, we evaluate the effectiveness of the HPAS framework through both synthetic and field data experiments. First, we assess the denoising performance and amplitude preservation capabilities of the HPAS framework on synthetic data. To precisely evaluate the denoising results, we quantify the denoised data using numerical metrics, frequency–wavenumber (F–K) spectral analysis, and single-trace waveform comparisons. Second, we apply the proposed method to a field ocean-bottom cable (OBC) dataset from the North Sea Volve field (Equinor, 2018) and compare its performance against polarization filtering and the Radon transform, thereby demonstrating the practical effectiveness of the HPAS framework in suppressing S-wave leakage.

## Synthetic examples

The synthetic data are generated by solving the acoustic wave equation via the finite-difference method. A horizontally layered velocity model is built, with P-wave velocities ranging from 1500 to 4500 m/s. A total of 1400 sources are uniformly distributed 5 m below the sea surface with a shot interval of 25 m, and 1400 receivers are located on the sea floor with a receiver interval of 25 m. A 16 Hz Ricker wavelet with a recording time of 8 s and a sampling interval of 0.004 s is used as the source signal. We apply the Radon transform to the horizontal component to extract the S-wave data, and estimate its incidence angle based on its correlation with the Z-component. The extracted S-wave data are then projected using the incidence angle to simulate the S-wave leakage, which is subsequently added to the synthetic P-wave data to generate the noisy data. The training noise is generated by performing moment matching between the S-wave data from the horizontal components with those of the simulated S-wave leakage derived from incident-angle-based processing.

TABLE 2

| | Radon transform | HPAS |
|---|---|---|
| SNR | 1.3423 | 11.7066 |
| SSIM | 0.6995 | 0.7626 |

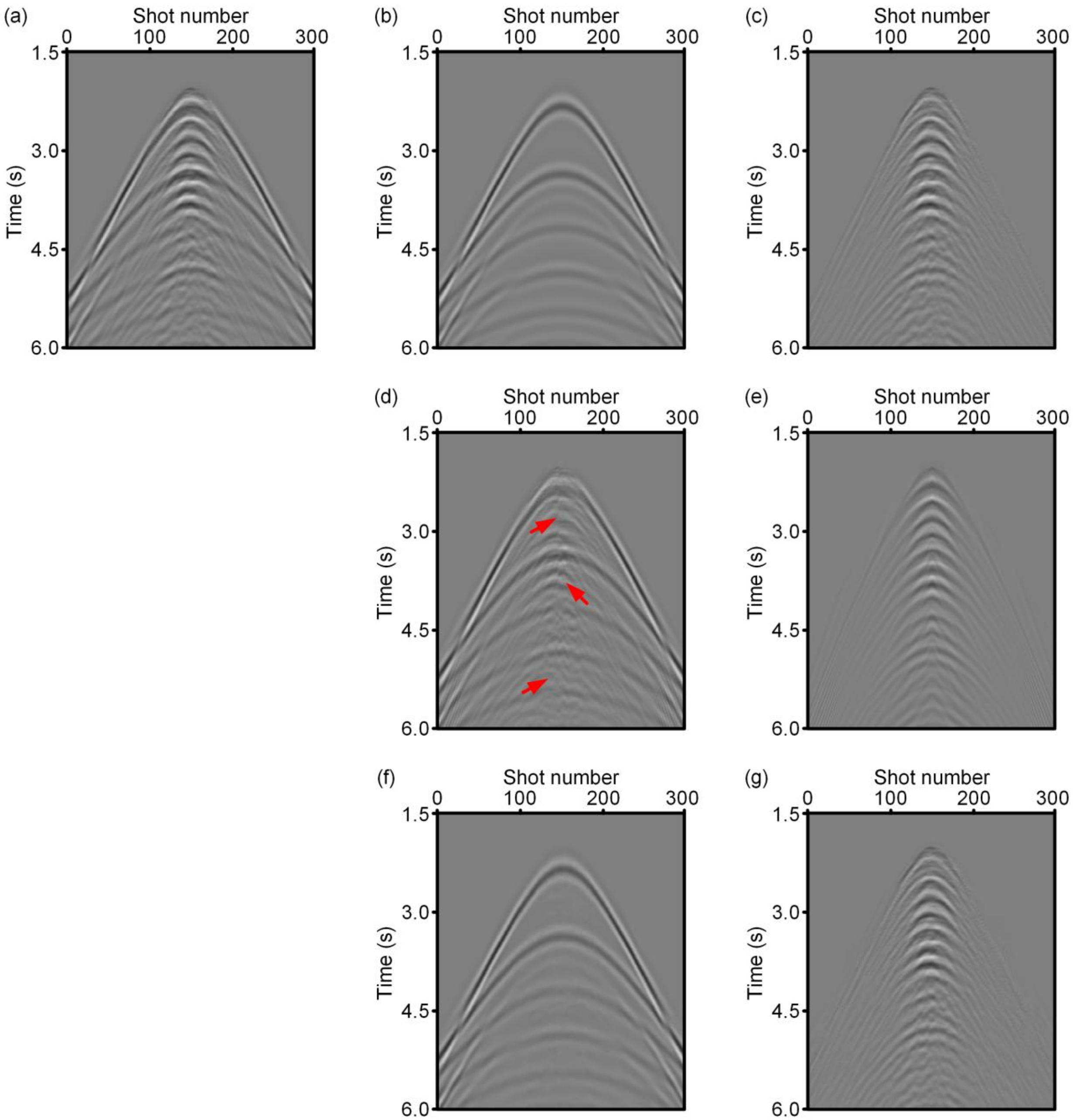

Figure 2. Denoising result of synthetic data with real S waves. (a) The noisy data. (b) The clean data. (c) The real S waves extracted from the horizontal components. (d) The denoised result of noisy data using the Radon transform. (e) Separated S waves using the Radon transform. (f) The denoised result of noisy data using the HPAS framework. (g) Separated S waves using the HPAS framework.

Figure 2a shows the noisy data. It can be observed that a portion of the early arrivals and reflections are contaminated by the S-wave leakage. Figures 2b and 2c show the clean P-wave data and the S-wave leakage, respectively. Figures 2d and 2e show the denoising results and the separated S-wave leakage via the Radon transform, respectively. Significant residual S-wave leakage remains in the denoising results of the Radon transform (highlighted by the red arrows). Conversely, Figures 2f and 2g show the predicted results and separated S-wave leakage via the HPAS framework. The prediction of the HPAS framework exhibits clear seismic events, with most of the S-wave leakage effectively suppressed. In addition, the separated S-wave leakage contains almost no residual P-wave data, which demonstrates that the proposed method is highly effective in suppressing S-wave leakage without damaging the P-wave signals. Table 2 quantitatively compares the performance of the Radon transform and the HPAS framework. The prediction of the HPAS framework achieves a significantly higher SNR than that of the Radon transform, and the corresponding SSIM values are closer to 1.

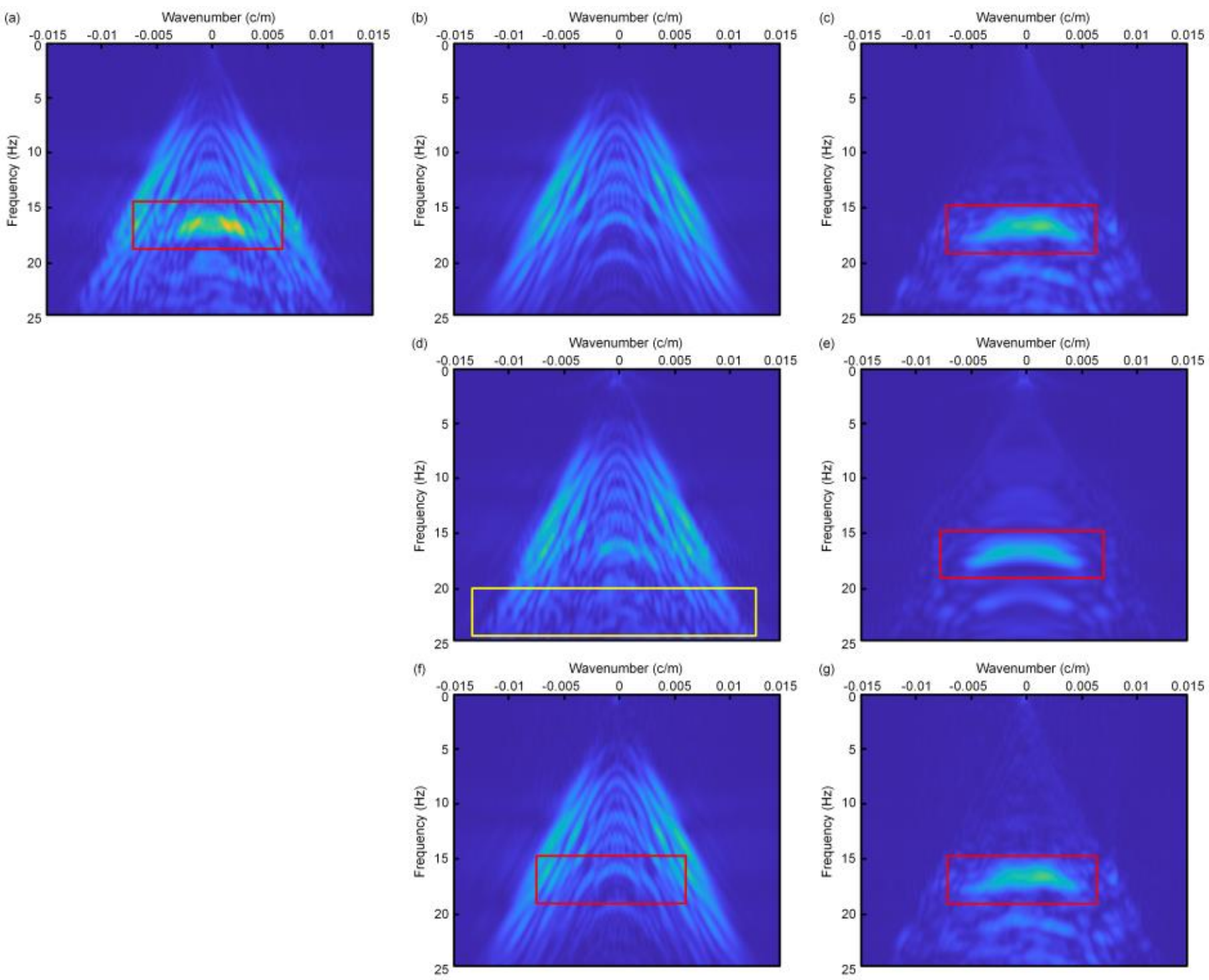

Figure 3. The F-K spectra of the denoising result of synthetic data with real S waves. (a) The noisy data. (b) The clean data. (c) The real S waves extracted from the horizontal components. (d) The denoised result of noisy data using the Radon transform. (e) Separated S waves using the Radon transform. (f) The denoised result of noisy data using the HPAS framework. (g) Separated S waves using the HPAS framework.

The F-K spectra of the aforementioned data are illustrated in Figure 3. Figures 3a, 3b, and 3c display the input data, the clean P-wave data, and the S-wave leakage, respectively. It can be observed that the F-K spectra of the clean P-wave data and S-wave leakage significantly overlap. The S-wave leakage contains substantial energy within the frequency band of 15–20 Hz, which affects the spectral characteristics of the input data (marked by the red box). Figures 3d and 3e show the prediction and the corresponding separated S-wave leakage obtained via the Radon transform, respectively. In the prediction of the Radon transform, significant residual S-wave leakage remains within the frequency band of 20–25 Hz (indicated by the yellow box). Furthermore, the energy of the S-wave leakage separated by the Radon transform is weaker than that of the ground truth (highlighted by the red box). Figure 3f presents the prediction of the HPAS framework, demonstrating the successful recovery of the P-wave signal. Importantly, the proposed HPAS framework achieves effective suppression of the S-wave leakage even in the most severely affected frequency bands (15–20 Hz, marked by the red box). Figure 3g presents the S-wave separated by the HPAS framework, which is highly consistent with the ground truth and exhibits almost no signal leakage (marked by the red box). These F-K spectra further demonstrate that the HPAS framework can effectively suppress S-wave leakage while preserving the P-wave characteristics in the frequency-wavenumber domain.

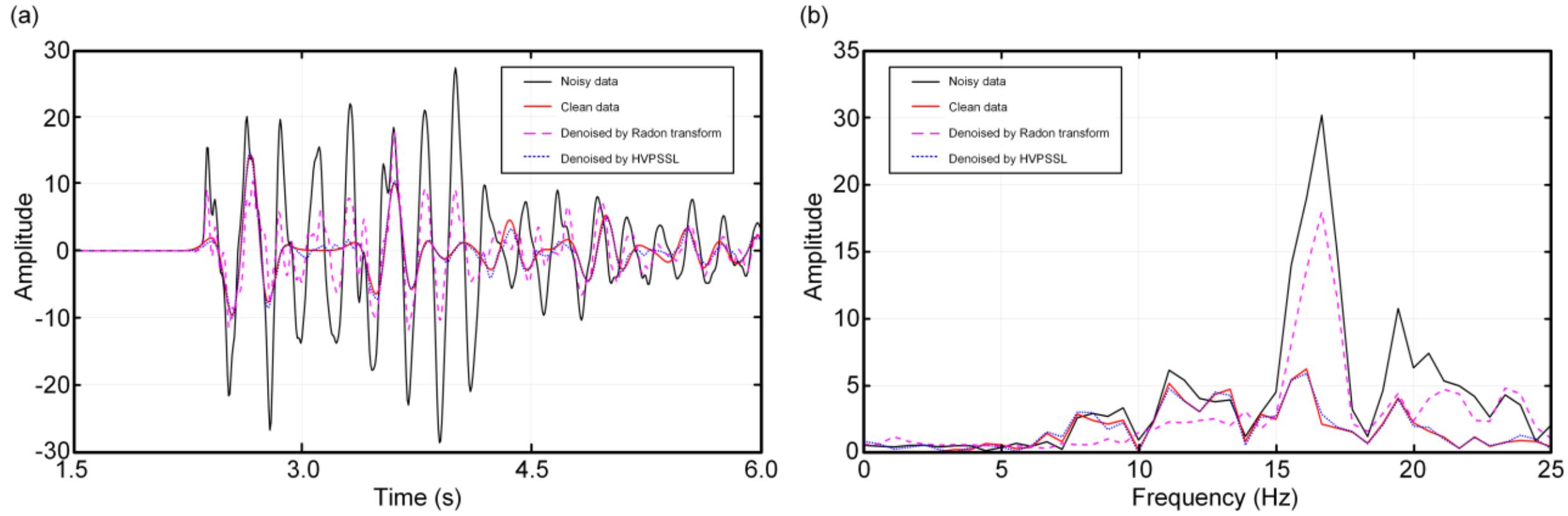

Figure 4. The comparison of the 150th shot of predictions. (a) The comparison of predictions in the time domain of 1.5–6.0 s. (b) The comparison of predictions in the frequency domain of 0–25 Hz.

Figure 4 illustrates the denoising performance for the 150th shot in a CRG. As shown in Figure 4a, strong S-wave leakage remains between 2.7 and 4.5 s, which severely distorts the P-wave signals. Evidently, the Radon transform fails to effectively suppress this intense S-wave leakage. In contrast, the proposed HPAS framework successfully attenuates the S-wave leakage, with the predicted phases closely aligning with those of the clean P-wave data. Figure 4b displays the corresponding amplitude spectra from 0 to 25 Hz. Due to the influence of S-wave leakage, a significant energy discrepancy exists between the noisy and clean data within the frequency band of 15–25 Hz. In this specific frequency band, the denoising result of the Radon transform deviates considerably from the ground truth. However, the proposed method effectively recovers the frequency characteristics of the P-wave signals, demonstrating its superior denoising performance.

**Field examples**

We further apply the proposed method to the field OBC data acquired from the North Sea Volve field. The field OBC data have a sampling interval of 0.004 s, with both receiver and shot intervals set to 25 m. To better demonstrate the comparison of denoising performance, the direct waves are muted. Figure 5a shows the P-component of the field data. Although the P- and Z-components correspond to different physical quantities, they record the P-wave responses from the same geological interfaces. Therefore, the valid P-wave signals in these two components exhibit similar waveforms. This property makes the P-component a useful reference for qualitatively validating the effectiveness of our method in preserving the P-wave signal, even though it remains entirely independent of the training process. Figure 5b shows the Z-component of the field data, which contains events similar to those in the P-component (marked by the yellow box), as well as the S-wave leakage (marked by red arrows). Figures 5c and 5f show the P and S waves separated from the Z-component using polarization filtering, respectively. It can be observed that the separated P waves obtained by polarization filtering still contain S-wave leakage (marked by red arrows). Moreover, the continuity of the P-wave events is poor (marked by the yellow box). In addition, residual signal energies are also present in the S-wave leakage separated by the polarization filtering (marked by the yellow box). Figures 5d and 5g show the P and S waves separated from the Z-component using the Radon transform, respectively. It can be observed that the P-wave separated by the Radon transform still exhibits S-wave leakage (marked by red arrows). Furthermore, there are noticeable residual P-wave signals in the S-wave separated by the Radon transform (marked with yellow boxes). Figure 5e shows the prediction of the HPAS framework. It can be observed that the HPAS framework effectively suppresses most of the S-wave leakage. The P-wave events in the prediction are continuous (marked by the yellow box) and exhibit strong similarities to the waveforms of the P-component. Figure 5h shows the separated S waves by the HPAS framework, which is highly similar to the S-wave leakage marked in Figure 5b (marked by red arrows). Therefore, the HPAS framework can effectively suppress S-wave leakage in the field OBC data and preserve the continuity of P-wave events.

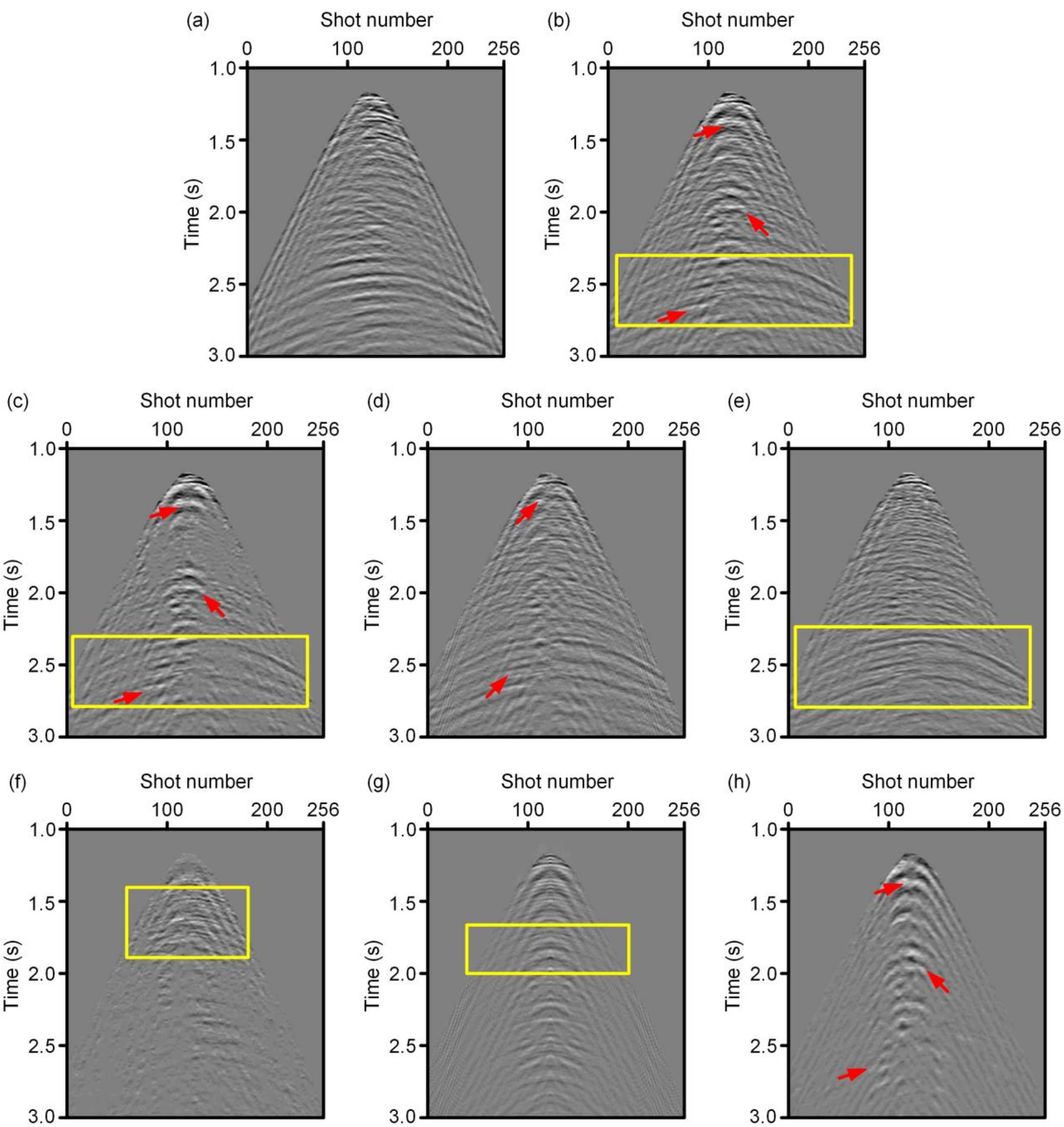

Figure 5. S-wave leakage suppression based on the field OBC data. (a) The P-component data. (b) The Z-component data. (c) The denoised result of Z-component data using polarization filtering. (d) The denoised result of the Z-component data using the Radon transform. (e) The denoised result of the HPAS framework. (f) Separated S waves using polarization filtering. (g) Separated S waves using the Radon transform. (h) Separated S waves using the HPAS framework.

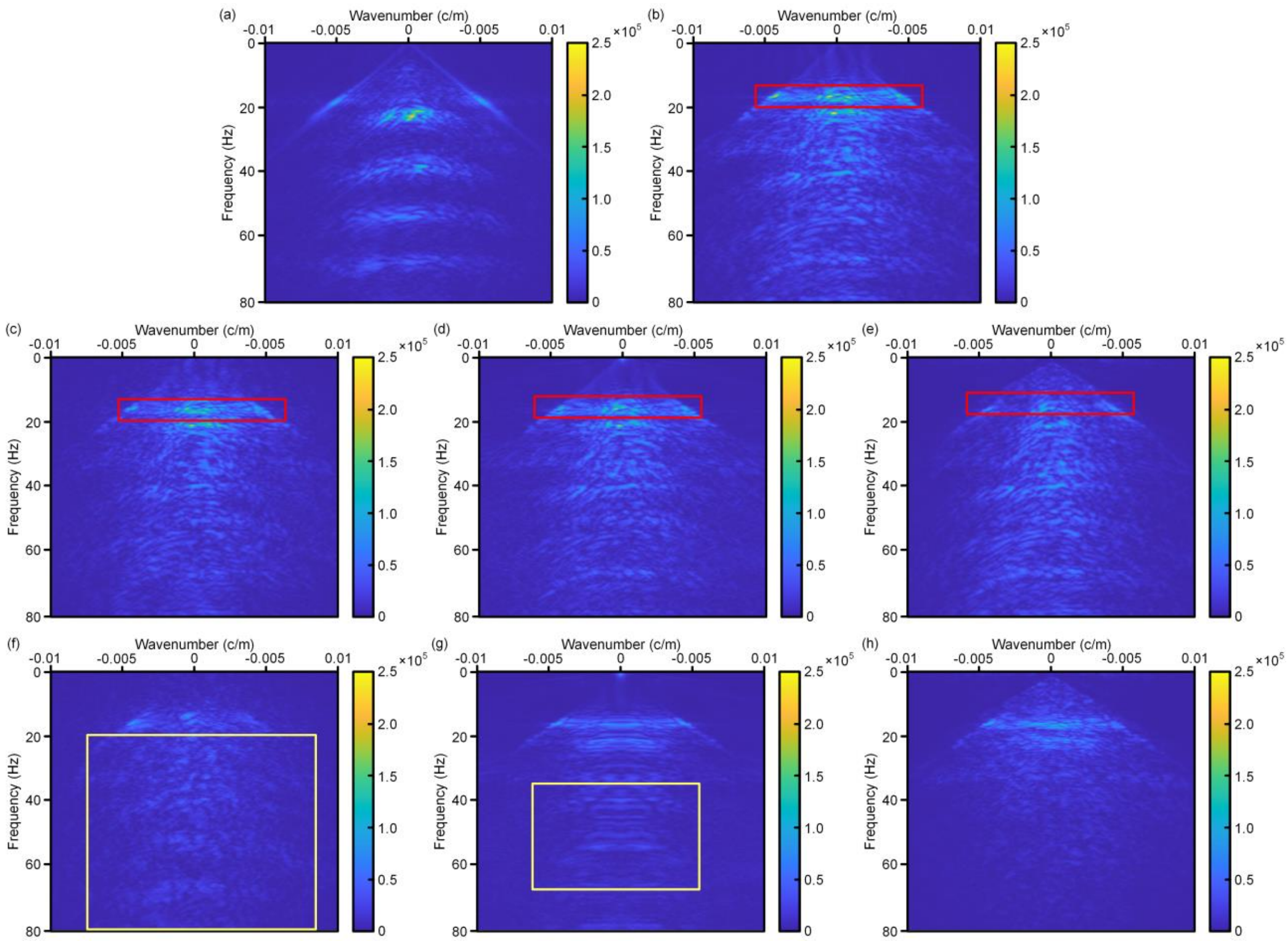

Figure 6. The F-K spectra of the S-wave leakage suppression based on the field OBC data. (a) The P-component data. (b) The Z-component data. (c) The denoised result of Z-component data using polarization filtering. (d) The denoised result of the Z-component data using the Radon transform. (e) The denoised result of the HPAS framework. (f) Separated S

waves using polarization filtering. (g) Separated S waves using the Radon transform. (h) Separated S waves using the HPAS framework.

We compare the F-K spectra of Figures 5a–5h in Figure 6. Figures 6a and 6b show the F-K spectra of the raw P- and Z-components, respectively. It can be observed that in Figure 6b, significant S-wave leakage exists in the frequency band of 12–20 Hz (marked by the red box). Figure 6c shows the P waves obtained by the polarization filtering, where the P and S waves between 12–20 Hz are not completely separated (marked by the red box). Figure 6f shows the S-wave leakage separated by the polarization filtering, where signal leakage is present between 20–80 Hz range (marked by the yellow box). Figures 6d and 6g show the P and S waves separated from the Z-component using the Radon transform. The P waves separated by the Radon transform exhibit noticeable shear-wave leakage between 12–20 Hz (marked by the red box). Signal distortion and signal leakage induced by the Radon transform can be observed in the S waves separated by the Radon transform (marked by the yellow box). Figure 6e shows the prediction of the proposed HPAS framework, which successfully achieves a better preservation of P-wave energy in the F-K domain and exhibits strong similarity to the P-component (marked by the red box). Figure 6h shows the S-wave leakage predicted by the HPAS framework, where no obvious P waves are observed between 12–20 Hz. Compared with traditional methods, the proposed HPAS framework effectively suppresses S-wave leakage and preserves P-wave signals, demonstrating its superiority in amplitude preservation.

## DISCUSSION

### Justification of Data Suitability for the HPAS Framework

TABLE 3

| | Separated noise | Training noise |
|---|---|---|
| Mean | $3.1228\times10^{-4}$ | $2.5134\times10^{-4}$ |

TABLE 4

| | Separated P waves and separated noise | Separated P waves and training noise |
|---|---|---|
| Correlation coefficient | 0.0280 | -0.0046 |

The proposed HPAS framework is based on two assumptions: the mean values of both the field S-wave leakage (**n**) and the simulated training noise (**s**) are zero, and the P waves in the Z-component are independent of both the S-wave leakage in the Z-component and the training noise. The mean values of the S-wave leakage separated by the Radon transform and the training noise are shown in Table 3. The mean values of the separated S waves and the training noise are $2.5134\times10^{-4}$ and $3.1228\times10^{-4}$, respectively. Both values are close to 0, validating the zero-mean assumption and demonstrating the suitability of the data for the proposed method. Furthermore, we use the mutual information (MI) (Duncan, 1970) to quantify the dependency between the P waves and S-wave leakage separated from the Z-component, as well as between the P waves separated from the Z-component and the training noise. The MI is defined as follows:

$$\mathrm{MI}(\mathbf{Z};\mathbf{G}) = \sum_{z\in\mathbf{Z}} \sum_{g\in\mathbf{G}} \mathrm{p}(z,g)\log\frac{\mathrm{p}(z,g)}{\mathrm{p}(z),\mathrm{p}(g)}, \tag{13}$$

where $\mathbf{Z}$ represents the P waves separated from the Z-component, and $\mathbf{G}$ represents either the S-wave leakage separated from the Z-component or the training noise. $z$ and $g$ represent the samples in $\mathbf{Z}$ and $\mathbf{G}$, respectively. $\mathrm{p}(z,g)$ represents the joint distribution of $z$ and $g$, while $\mathrm{p}(z)$ and $\mathrm{p}(g)$ represent the marginal distributions of $z$ and $g$, respectively. In general, the MI value approaching zero indicates a higher degree of statistical independence. As shown in Table 4, the MI between the P waves and the S-wave leakage separated from the Z-component is 0.0261, and the MI between the P waves and the training noise is 0.0735. These low MI values indicate that the data used for denoising are approximately independent, thereby supporting the independence assumption required by the proposed HPAS framework.

## Generalization analysis

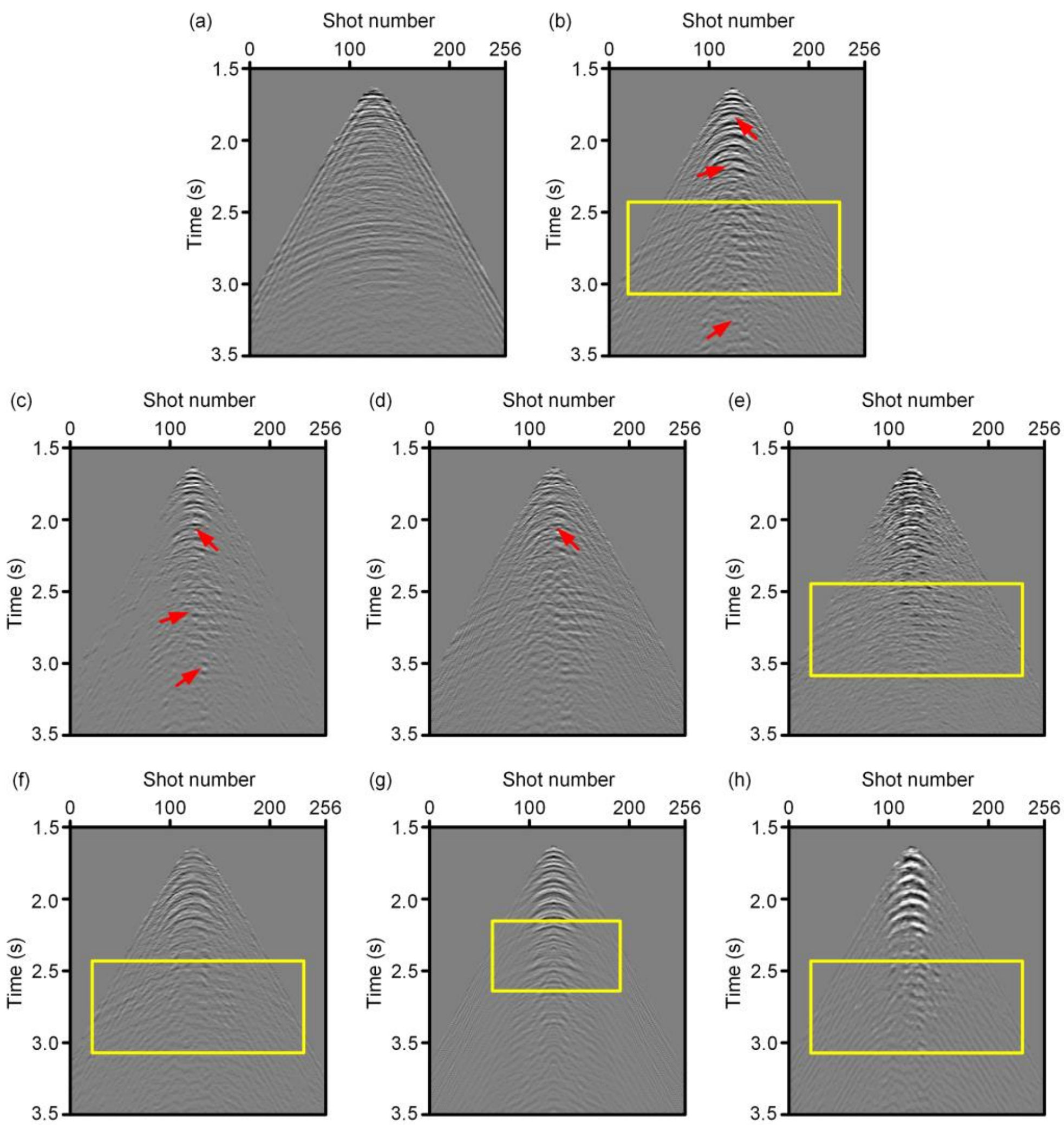

Figure 7. S-wave leakage suppression based on the field OBC data from another survey line. (a) The P-component data. (b) The Z-component data. (c) The denoised result of Z-component data using polarization filtering. (d) The denoised result of the Z-component data using the Radon transform. (e) The denoised result of the HPAS framework. (f) Separated S waves using polarization filtering. (g) Separated S waves using the Radon transform. (h) Separated S waves using the HPAS framework.

To verify the generalization ability of the proposed HPAS framework, we apply it to the Z-component of OBC data acquired along another survey line from the North Sea Volve field. This survey line is located 1.6 km away from the line corresponding to the data shown in Figures 5a and 5b. Figures 7a and 7b show the P-and Z-components of the OBC data, respectively. The P-wave signal and S-wave leakage in the Z-component are marked by yellow boxes and red arrows, respectively. Figures 7c and 7f show the P and S waves separated by polarization filtering, respectively. It can be observed that the separated P waves contain noticeable residual S-wave leakage (marked by red arrows). Meanwhile, the separated S waves exhibit obvious missing samples in the high-amplitude regions (marked by the yellow box). Figures 7d and 7g show the denoised results and the separated S-wave leakage from the Radon transform, respectively. Some shear-wave leakage remains in the denoised results via the Radon transform (marked by the red arrow). Additionally, leaked P-wave energies can be observed in the separated S-wave leakage (marked by the yellow box). Figure 7e shows the prediction of the proposed method, where the recovered P waves exhibit better continuity (marked by the yellow box). Figure 7h shows the S-wave leakage removed by the proposed method. The high-amplitude S-wave leakage is effectively suppressed, and the separated S-wave leakage by the proposed method is more continuous than that obtained by traditional methods (marked by the yellow box). Overall, the proposed method achieves satisfactory accuracy on another survey line, which demonstrates the generalization.

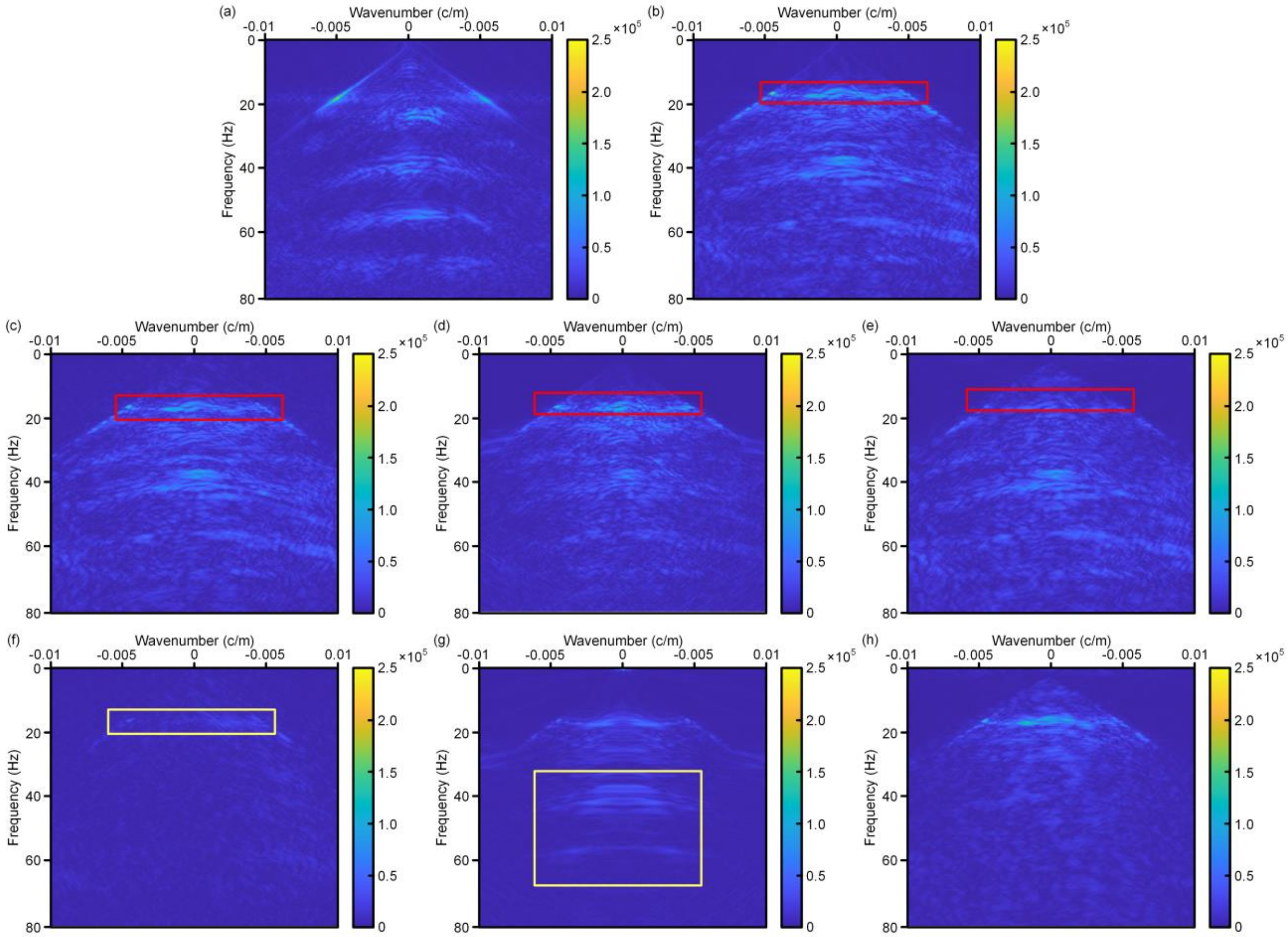

Figure 8. The F-K spectra of the S-wave leakage suppression based on the field OBC data from another survey line. (a) The P-component data. (b) The Z-component data. (c) The denoised result of Z-component data using polarization filtering. (d) The denoised result of the Z-component data using the Radon transform. (e) The denoised result of the HPAS framework. (f) Separated S waves using polarization filtering. (g) Separated S waves using the Radon transform. (h) Separated S waves using the HPAS framework.

Figure 8 shows the F-K spectra of the corresponding data shown in Figure 7. Figures 8a and 8b show the F-K spectra of the P- and Z-components, respectively. Significant energies of the S-wave leakage can be observed between 12–20 Hz in the F-K spectra of the Z-component (marked by the red box). Figures 8c and 8f show the F-K spectra of the P and S waves obtained by polarization filtering, respectively. In Figure 8c, significant S-wave energy remains in the 12–20 Hz band and wavenumber range of −0.005 to 0.005 (marked by the red box). In contrast, only weak S-wave energy is observed in the corresponding region of Figure 8f (marked by the yellow box), indicating that polarization filtering fails to effectively separate P and S waves. Figures 8d and 8g illustrate the denoised results and the separated S-wave leakage obtained by the Radon transform, respectively. Figures 8d and 8g show the denoised result and the separated S-wave leakage obtained by the Radon transform, respectively. In Figure 8d, residual energy remains within the 12–20 Hz frequency band and wavenumber range of −0.005 to 0.005 (marked by the red box). In Figure 8g, energy distortion induced by the Radon transform is observed in the 30–70 Hz band within the same wavenumber range (−0.005 to 0.005; marked by the yellow box), indicating limitations in both noise attenuation and signal preservation. Figure 8e shows the denoised result of the proposed method, where the S-wave energy in the 12–20 Hz frequency band and wavenumber range of −0.005 to 0.005 (marked by the red box) is effectively suppressed. Figure 8h shows the separated S-wave leakage, in which the S-wave energy within the same frequency and wavenumber ranges is largely captured. These results demonstrate that the proposed method maintains its denoising performance across different survey lines.

## CONCLUSION

To address the issue of S-wave leakage in the Z-component of OBC data, we propose an adaptive shear-wave leakage suppression framework based on horizontal-component priors. Specifically, we use the Radon transform and moment-matching to generate training noise from horizontal components, which is statistically consistent with the S-wave leakage. The input data and pseudo-labels are then constructed by adding and subtracting the training noise to the original Z-component. Theoretical derivations and data validations demonstrate that the proposed method can establish a complex mapping between noisy and clean data. Compared with conventional methods, the proposed HPAS framework exhibits superior denoising accuracy and generalization capability. In summary, the proposed framework provides a valuable reference for designing other deep-learning-based noise suppression methods. In our future work, we plan to explore more efficient network architectures to further enhance denoising performance. In summary, the proposed HPAS framework presents a robust, label-free approach for S-wave leakage suppression in multi-component seismic data processing. Future research will focus on improving the network architecture to further enhance the denoising performance on more complex data.

## ACKNOWLEDGMENTS

This work was supported by the National Natural Science Foundation of China (no. 42574169) and the National Science and Technology Major Project for Deep Earth Exploration (no. 2025ZD1007603).

## DATA AND MATERIALS AVAILABILITY

Data associated with this research are available and can be obtained by contacting the corresponding author.